\documentclass[12pt]{article}
\usepackage{graphicx} 

\usepackage{authblk}
\usepackage{xurl}
\usepackage{amsmath}
\usepackage{hyphenat}

\title{When Common Law Ages: Two Centuries of Growing Inertia in US Judicial Opinions}

\author[1,2,*]{Seoul Lee}
\author[3,*]{Taekyun Kim}
\author[4]{Jisung Yoon}
\author[5,6,**]{Hyejin Youn}

\affil[1]{Kellogg School of Management, Northwestern University, Evanston, IL, US}
\affil[2]{Northwestern Institute on Complex Systems, Evanston, IL, US}
\affil[3]{School of Business, Chungnam National University, Daejeon, Korea}
\affil[4]{KDI School of Public Policy and Management, Sejong, Korea}
\affil[5]{College of Business Administration, Seoul National University, Seoul, Korea}
\affil[6]{Santa Fe Institute, Santa Fe, NM, US}
\affil[*]{S.L. and T.K. contributed equally to this work.}
\affil[**]{Correspondence can be sent to h.youn@snu.ac.kr}

\date{\today}

\begin{document}

\maketitle

\begin{abstract}
Judicial opinions once considered sound can lose relevance over time. Yet, little has been known, both systematically and at scale, about how judicial reasoning has evolved. Here, we analyze four million US court decisions from 1800 to 2000, quantifying each ruling’s \emph{disruptiveness}—how much it breaks from established citation pathways. We find that such pathbreaks have declined over time, indicating that courts have become increasingly constrained by precedent. This growing inertia appears to be driven by two structural factors. The first is precedent overload, evidenced by the volume of case law outpacing population growth (scaling exponent $\approx1.7$). The second is the rise of ideological polarization within the judiciary, which introduces institutional uncertainty that prompts greater deference to established precedent. Despite this overall tendency toward path dependence, we find that a relatively small number of high-authority courts continue to shape legal discourse through top-down interventions. Our findings recast legal reasoning as an evolutionary process shaped by structural growth, institutional memory, and hierarchical structure, incorporating broader theories of innovation and organizational adaptation into the study of law.
\end{abstract}

\section*{Introduction}

On a spring day in 1954, the US Supreme Court issued a decision that struck down racial segregation in public schools, overturning nearly six decades of precedent set by \textit{Plessy v. Ferguson} (1896). This landmark decision in \textit{Brown v. Board of Education} (1954) rippled through classrooms, courthouses, and communities across the nation, marking a pivotal moment when the law responded to society's demands for justice, dignity, and institutional change. \textit{Brown} was not an anomaly in history. In \textit{Strauder v. West Virginia} (1880), the Court invalidated the exclusion of Black jurors. Nearly a century later, \textit{Reed v. Reed} (1971) ruled against automatic gender preference in estate administration. These cases remind us that courts can do more than just follow precedent—they can break and disrupt it.   

Law, in this view, is not an immutable set of rules but a complex adaptive system that evolves in response to social feedback \cite{Elliott1985, katz2020complex, vivo2024complexity, Koehler2022, Rockmore2018, Fernandes2025}. In particular, a common law system evolves through gradual, case-by-case reasoning. Courts achieve this by both relying on precedent as institutional memory and adjusting it to novel contexts \cite{Elliott1985, Niblett2010}. However, this adaptive capacity is not limitless. As precedent accumulates, the legal system risks “overfitting” to its own history, making its structure increasingly interdependent and complex \cite{li2015law, katz2020complex}. Furthermore, the mounting burden of precedent raises the cost of legal coordination. That is, the increasing volume of the legal system reduces its ability to reorganize in response to societal change, thereby making the system more rigid \cite{Koehler2022, march2000dynamics}. In this sense, precedent—originally a tool for learning and flexibility—can become a structural constraint, generating inertia that impedes legal evolution \cite{march2000dynamics}. 

Structural inertia is not a new problem. It is often a natural consequence of institutional aging. As organizations mature, much like humans, they tend to resist change, even when external conditions shift \cite{hannan1984structural}. Over time, rules and procedures become formalized \cite{yoon2023individual}, reinforcing commitments to precedent and internal coherence \cite{rheinstein1954max, nonet2017law}. When such commitments dominate, legal systems risk entering a “crisis of formality,” in which responsiveness to social change is constrained \cite{teubner1983substantive, nonet2017law}. Increasing political polarization in recent times has even intensified this inertia, creating legislative gridlock and eroding the consensus necessary for reform \cite{binder1999dynamics, mayhew1991divided, thurber2015american}.

Nevertheless, the demand for adaptation has never been greater. Legal systems now operate in a world of rapid technological change and growing institutional and cultural complexity. Under such conditions, continuity alone may no longer suffice. This raises a crucial research question: Do courts still deviate from precedent to address evolving societal needs as they once did, or have they become caught in the iron cage of precedent? 

Here, we address this question empirically. Analyzing 3,976,542 judicial opinions issued between 1800 and 2000, we measure whether a given ruling reinforces existing legal pathways or breaks from them \cite{funk2017dynamic, park2023papers}. We use legal citation to measure these interpretive dynamics---how precedent is maintained, reinterpreted, and overturned in judicial reasoning \cite{spriggs2000measuring}.

We find that over the past two centuries, courts have become increasingly inclined to follow established legal pathways, consistent with the broader pattern of legislative gridlock and institutional rigidity in politics \cite{binder1999dynamics, mayhew1991divided, thurber2015american, Pocheptsova2009}. This growing reliance on precedent appears to be driven by two structural factors. First, the volume of case law has expanded far more rapidly than the population, scaling superlinearly with an exponent of approximately 1.7. This runaway growth imposes increasing cognitive and institutional burdens, making it harder for judges to reorganize or reinterpret the legal landscape. In essence, as the mass of precedent accumulates, so does its inertia, reducing the system’s capacity to shift direction. Second, rising ideological polarization within the judiciary has increased institutional uncertainty, incentivizing judges to rely more heavily on the interpretive stability of \textit{stare decisis} \cite{Pocheptsova2009}. As it becomes harder to reach consensus, precedent becomes not only a tool for consistency, but also a defense against ambiguity.

Still, inertia does not govern all periods. We identify two conditions under which doctrinal disruption becomes more likely. First, major political realignments, such as the Reconstruction era following the Civil War, introduce external shocks that open space for legal reorganization. Second, a relatively small number of high-authority courts, particularly the US Supreme Court, continue to exert top-down influence, redirecting legal trajectories through landmark rulings, as in \textit{Strauder} (1879), \textit{Brown} (1954), and \textit{Dobbs v. Jackson Women’s Health Organization} (2022).

Taken together, these findings suggest that the evolution of legal systems can be understood as a complex adaptive process. Legal systems are shaped by the interplay between scale, memory, and authority. Under ordinary conditions, these features stabilize doctrine. Only under external pressures, such as political realignments and shifts in interpretive authority, can they create conditions for change. Legal disruption, in this view, is not a mere breakdown of precedent but a structural reorganization of interpretive pathways. As such, our findings recast legal reasoning as a dynamic process shaped by structural scale, institutional memory, and hierarchical structure, linking the evolution of law to broader patterns in innovation, cultural change, and organizational adaptation.

\section*{Measuring disruption in case law}

\begin{figure*}[!t]
    \centering
    \includegraphics[width=0.99\textwidth]{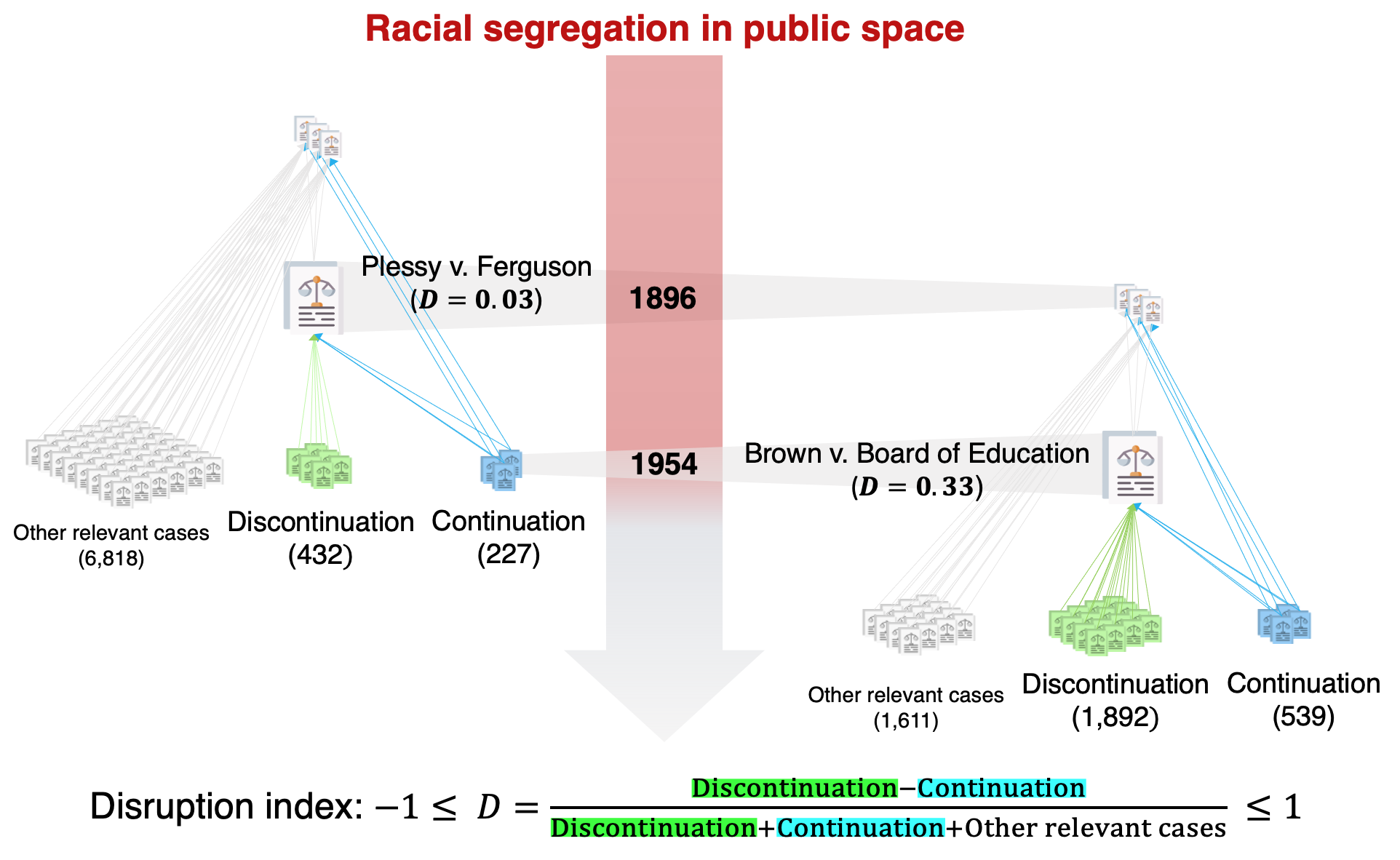}
    \caption{\textbf{Schematic Illustration of Aligned vs. Misaligned Cases in the Legal Doctrine of Racial Segregation}. Cases are arranged chronologically from top to bottom, spanning two centuries (1800–2020). Two landmark Supreme Court decisions, \textit{Plessy v. Ferguson} (1896) and \textit{Brown v. Board of Education} (1954), are selected as focal points to illustrate how legal citation paths can either align with or diverge from existing precedent. For each focal case $i$, the disruption index $D_i$ measures the relative difference between subsequent cases that cite only the focal case (green) and those that cite both the focal case and its predecessors (blue). In this way, a higher $D$ value, approaching 1, indicates greater misalignment with the precedential structure, meaning the focal case represents a doctrinal break rather than a continuation. Note that disruption is not assessed from the textual content of a ruling but its structural position in the evolving network of legal citations—how future cases position the focal case within the broader legal reasoning. For example, 34\% of cases citing \textit{Plessy} also cite its predecessors, compared to only 22\% for \textit{Brown}. As a result, \textit{Brown} registers a substantially higher disruption score ($D_{\textit{Brown}} = 0.33$) than \textit{Plessy} ($D_{\textit{Plessy}} = 0.03$), consistent with its widely recognized role in overturning the legal foundations of racial segregation.
 }
    \label{fig:schematic}
\end{figure*}

In the United States, published judicial opinions collectively constitute case law. These opinions are considered law because of their binding authority for future rulings on similar legal issues. This principle of precedential power, known as \textit{stare decisis}, ensures consistency and stability in the judicial process over time. Yet, within this framework, judges still have the discretion to reinterpret or adapt established legal principles in response to evolving societal needs. Unlike statutory law, which requires formal legislative amendments to change, case law therefore evolves organically through judicial decisions that challenge or disrupt existing precedents.

Given the interconnected nature of judicial opinions through citation relationships, we conceptualize disruption as a structural change within the citation network of judicial opinions \cite{spriggs2000measuring}. A disruptive opinion breaks established pathways of citation, rendering its precedents less relevant for future legal reasoning. It is important to distinguish \emph{disruption} from \emph{novelty}, a related concept in the literature on creativity and innovation \cite{leibel2024we}. Novelty refers to the introduction of new, surprising, or atypical combinations of ideas \cite{uzzi2013atypical, Kim2016, youn2015invention, kwon2024}, but does not necessarily alter the trajectory of future work. Disruption, on the contrary, captures transformative shifts that redirect the development of a field.

Although citation-based measures of disruption were first formalized to study change in science and technology \cite{funk2017dynamic, wu2019large, park2023papers, leibel2024we, li2024breaking}, their intellectual lineage traces back to law. Shepard’s Citations, introduced in the 19th century, pioneered the idea of tracking interpretive influence across rulings, laying the groundwork for how citation structures are now used to study institutional evolution and innovation in law \cite{spriggs2000measuring, fowler2008, whalen2016legal, whalen2017common}. 

Parallels between scientific research and legal reasoning further support the use of such a citation-based approach in studying the evolution of case law. Both domains are cumulative systems composed of individual contributions (i.e., academic papers and judicial opinions, respectively), produced by different authors or judges over time. These individual works are interconnected in that each new work builds upon prior discoveries or precedents. Most contributions tend to reinforce existing frameworks, but some disrupt them by introducing transformative ideas, such as relativity theory and quantum mechanics in early 20th-century physics \cite{kuhn1997structure} and \textit{Strauder} and \textit{Brown} in US case law. These disruptions drive the evolution of their respective systems, advancing toward a deeper understanding of the physical world or substantive justice.

The analogy between scientific discovery and judicial reasoning indeed has a long intellectual history. Legal scholars have often described the judicial process as a kind of ``discovery" process \cite{hayek2022law, christainsen1989law}. Especially within the common law tradition, judges dealing with individual cases are seen as participants in a collective endeavor to discover what the law ought to be, though debates persist over the object of discovery, whether it be natural law \cite{fuller1965morality, finnis2011natural}, economic efficiency \cite{posner2014economic}, or other concepts of justice. This resembles the scientific quest to uncover laws of nature or universal patterns. In both domains, collective efforts leave traceable records in citation networks. Just as scientists cite prior studies to acknowledge intellectual debts, judges cite precedents to ground their rulings in established principles \cite{posner2000economic, whalen2017common}.

Given these behavioral and structural similarities between science and case law, we apply the citation-based index of disruption developed for science and inventions \cite{funk2017dynamic} to measure the disruptiveness of judicial decisions. The disruption index, $D$, calculates the normalized difference between two counts: the number of future decisions citing the focal decision but none of its references and the number of future decisions citing both the focal decision and its references. A higher value of $D$ indicates that the focal decision has shifted attention away from prior decisions or existing citation pathways. In essence, disruptive decisions establish new legal guidelines that deviate from established precedents, creating new chains of legal precedents. In the following, to illustrate the application of $D$ in case law, we analyze two landmark cases on racial segregation: \textit{Plessy v. Ferguson} (1896) and \textit{Brown v. Board of Education} (1954)

Government-led discriminatory practices against black people persisted in the US even after the official abolition of slavery. One such practice was racial segregation, which restricted black people’s access to public facilities. \textit{Plessy} (1896) further solidified this segregation by establishing the ``separate but equal” doctrine. This legal principle asserted that racial exclusion from public facilities was permissible as long as separate facilities were made available for each race. However, in 1954, the landmark decision of \textit{Brown} overturned the doctrine from \textit{Plessy} by ruling racial segregation in public schools unconstitutional. This decision disrupted the existing legal principle that had permitted segregation in public facilities. \textit{Brown} is now considered as a significant step toward ending racial segregation in the US.

Figure \ref{fig:schematic} illustrates the different impacts of these two cases on citation pathways within US case law. As \textit{Plessy} upheld the existing practice of racial segregation, subsequent judges handling relevant cases continued to cite both \textit{Plessy} and the earlier precedents on which it relied, maintaining established citation pathways. In contrast, \textit{Brown} rendered prior rulings that upheld racial segregation obsolete. Therefore, judges addressing similar cases after \textit{Brown} found relatively lttle need to reference pre-\textit{Brown} cases, as those cases no longer aligned with the new standard of interpreting the Constitution.

The disruption index, designed to capture this structural shift, is defined as:
\begin{equation}
    D = \frac{\text{discontinuation} - \text{continuation}}{\text{relevant future cases}} 
    \label{eq:d_index}
\end{equation}

Conceptually, $D$ quantifies the extent to which a case, relative to its predecessors, influences the direction of future legal decisions (see \textbf{Methods} for details). A higher proportion of future cases that cite only the focal case without citing its predecessors indicates a greater disruption of legal continuity, reflecting a break in the established line of doctrinal development. Note that disruption is not assessed from the textual content of the decision itself but from its structural position within the evolving network of legal citations. That is, it is assessed based on how future rulings incorporate or bypass the decision in constructing legal arguments. Within this framework, \textit{Brown} exhibits a disruption score of $D_{\textit{Brown}} = 0.33$, nearly ten times higher than \textit{Plessy}'s $D_{\textit{Plessy}} = 0.03$, consistent with its widely recognized role in reshaping the course of US constitutional jurisprudence.  

\section*{Growing inertia in case law}

\begin{figure*}[h!]
    \centering
    \includegraphics[width=.99\textwidth]{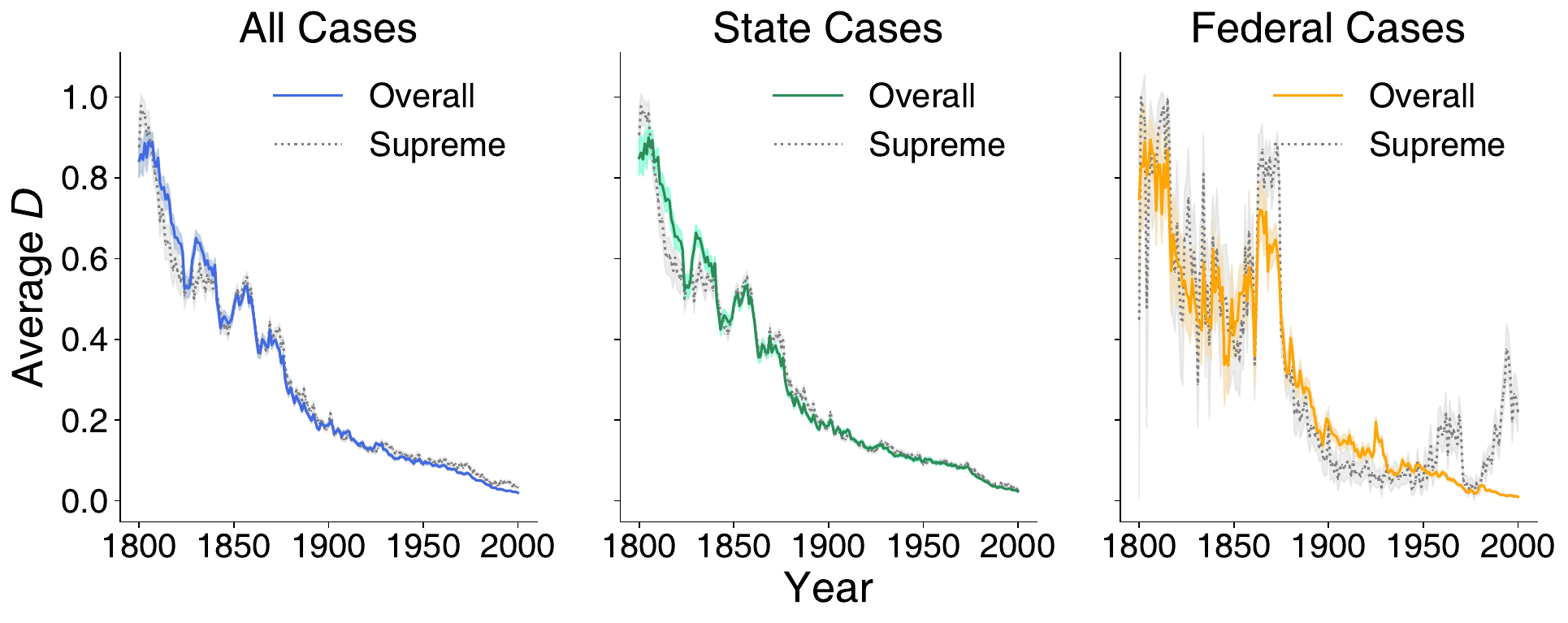}
    \caption{\textbf{Decrease in disruptive case law} The average disruptiveness of all US cases (N = 3,976,542) decreases over time. State cases (N = 3,173,146) follow a similar trend, while federal cases (N = 803,396) show a brief surge in disruptive decisions around the 1870s before joining the overall downward trend. The dotted lines indicate cases issued by the supreme courts at each level. While the disruptiveness of cases from state supreme courts consistently declines in line with the overall trend, cases from the federal supreme court show increases in disruptiveness around the 1960s and 2000s. Shaded bands represent the 95\% confidence interval around the mean. }
    \label{fig:d_by_year}
\end{figure*}

The sharp contrast in $D$ between \textit{Plessy} (1896) and \textit{Brown} (1954) shows that \textit{Brown} marked a much more significant break from the existing precedent structure. But how common is this kind of disruption? Are landmark ``breakers" like \textit{Brown} rare exceptions, or do they occur more regularly over time? To examine long-term trends in legal disruption, we calculate the disruption index of all 3,976,542 judicial opinions published in the US between 1800 and 2000. Figure~\ref{fig:d_by_year} plots the annual average of these values by year of decision. 

Our analysis shows that \textit{Brown} was indeed exceptional for its time. Most of the decisions in that period show $<\,D (t=1954)\,> \, \simeq 0.1$, which is far lower than $D_{\textit{Brown}}=0.33$. More broadly, we find that today's legal decisions are increasingly not breakers but builders of the existing precedent structure. This decline in legal disruption was especially sharp during the 19th century and began to level off throughout the 20th. The trend toward legal inertia holds consistently across jurisdictions, both state and federal.
 
Within this broader pattern of decline, a few transient peaks appear in federal case law. The most pronounced of these appears in the late 1800s, likely reflecting federal courts' efforts to help reconstruct the nation following the Civil War \cite{currie1984constitution}. This era was characterized by sweeping political, economic, and social transformation \cite{Kim1995}, during which federal courts were tasked with addressing major constitutional issues, particularly those involving the rights of newly emancipated slaves. Many decisions during this period broke from established precedent. For example, \textit{Strauder v. West Virginia} (1879) held that excluding black people from juries solely on the basis of race violated the Equal Protection Clause. Understandably, $D_{Strauder}=0.36$ is comparable to that of \textit{Brown}. Given that the Reconstruction era demanded substantial departures from prior legal doctrines, this brief surge in disruptiveness during the 1870s underscores the interpretive value of the disruption index as a measure of legal change.

Furthermore, the late 1800s observed an extraordinary wave of technological innovation and patenting activity \cite{youn2015invention, smil2005creating}. The introduction of telephones (1876), phonographs (1878), and typewriters (1868) triggered major social and economic shifts, prompting legal adaptations, especially in patent law, which is governed at the federal level. These developments likely contributed to structural changes in legal reasoning, as reflected in elevated disruption scores for related cases.  

Since supreme courts have the final say within their respective jurisdictions, we separately analyzed the average disruptiveness of judicial opinions issued by state supreme courts and the US Supreme Court. While state supreme courts have followed the overall trend of issuing increasingly less disruptive decisions, the US Supreme Court exhibited several surges in disruptiveness during the late 20th century. This pattern suggests that court hierarchy has a role in shaping judicial decisions' disruptiveness, which we will discuss in more detail in a following section.

In addition, we explore several variations of citation-based disruption measures to test the robustness of our analysis. Across all specifications, the declining trend in disruptiveness remains consistent (see Figures S1 and S2). These variants include adjustments to the citation window and the use of $D_{\text{nok}}$ \cite{bornmann2020disruption, wu2019solo}, accounting for the potential impact of citation inflation \cite{park2023papers}. In all cases, the overall downward trend persisted. 

\section*{Precedent Overload and Political Polarization}

\begin{figure*}[h!]
    \centering
    \includegraphics[width=.99\textwidth]{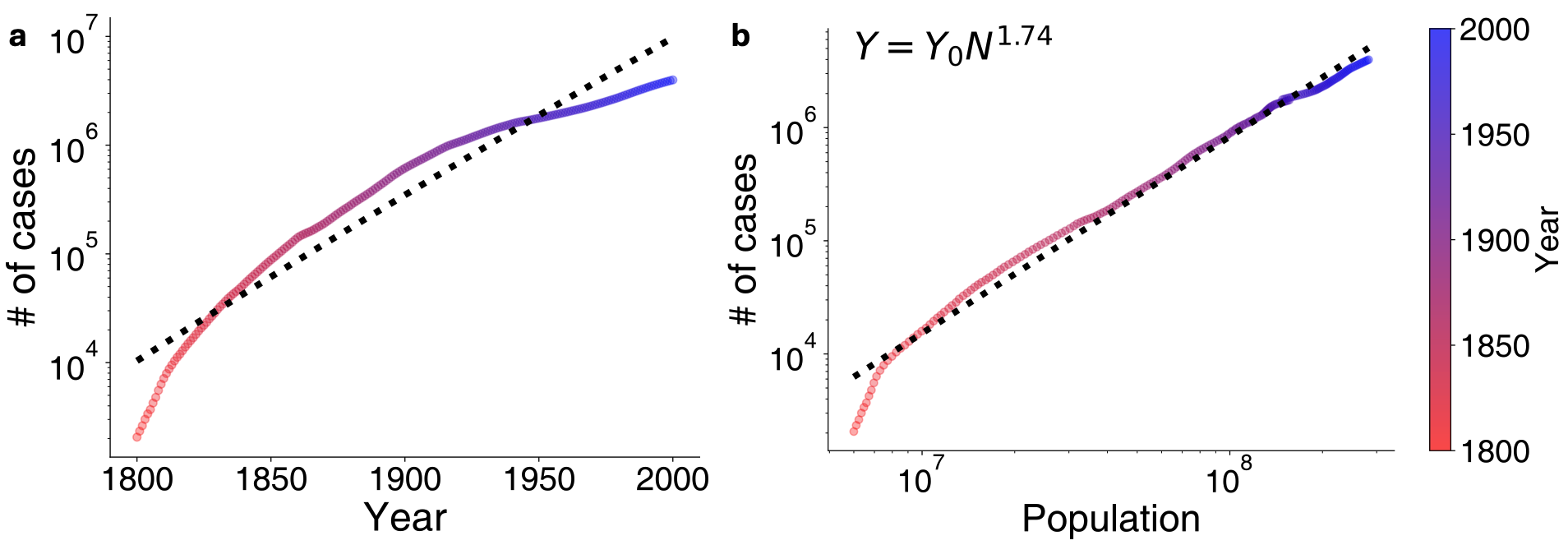}
    \caption{\textbf{Cumulative number of published legal cases over time} \textbf{a} growth by calendar year (nominal time) and \textbf{b} growth by population (event time). The maker's color indicates publication year, ranging from red (1800) to blue (2000). Notably, legal cases scale superlinearly with population ($Y \sim N^{\beta}$), with $\beta = 1.74$, as shown by the fitted black dotted line.}
    \label{fig:no_case_by_population.png}
\end{figure*}

What accounts for the long-term decline in legal disruption? We investigate two macro-level forces that may contribute to this increasing legal inertia. First, we examine the effect of legal proliferation---a defining feature of modern legal systems \cite{teubner1983substantive, nonet2017law}---on the rigidity of precedent. Second, we examine whether the rise of ideological polarization, which has been linked to institutional gridlock in legislative and administrative contexts \cite{binder1999dynamics, mayhew1991divided, thurber2015american}, also contributes to the growing inertia in the judicial domain by deterring lower courts from issuing disruptive rulings.

As Max Weber discusses, modern law tends to proliferate over time \cite{jennings2005weber, weber1978economy}, regulating increasingly diverse aspects of social life which had previously governed by informal norms \cite{katz2020}. Such an inflation in law can be effectively captured using the scaling framework, which seeks to explain system-level quantities as functions of system size (see \textbf{Methods}) \cite{bettencourt2007growth, youn2016scaling, west2018scale, west1997general, ash2024scaling}. The growth of case law may appear to have slowed when plotted simply against linear time (\ref{fig:no_case_by_population.png}\textbf{a}). However, when the volume is plotted against population size following the scaling framework, (Figure \ref{fig:no_case_by_population.png}\textbf{b}), it reveals a consistent growth rate relative to population, fitting well to a power law function $Y\sim N^\beta$ (black dashed line). The estimated exponent, $\beta=1.74$, exceeds $1$, which indicates that case law has been consistently expanding at a rate significantly higher than population growth. This superlinear growth aligns with patterns observed by prior research in many social systems \cite{youn2016scaling, Gomez2012}.

We propose that this rapid expansion of case law creates inertia, making it increasingly challenging for any single decision to disrupt the entire legal system. A parallel phenomenon occurs in science, where an overwhelming volume of publications leads researchers to favor established works over novel ideas \cite{chu2021slowed}. Similarly, judges, burdened by the sheer volume of case law \cite{Koehler2022}, may struggle to engage with disruptive ideas, opting instead to reinforce precedents. This cognitive overload is exacerbated by the interconnected nature of case law, where altering one precedent can ripple across different legal principles, prompting judges to weigh broader implications. Thus, we propose that the cumulative number of cases within a jurisdiction in a given year is negatively associated with the disruptiveness of its judicial decisions in that year.

Next, we investigate the role of political ideology in the decrease in disruption. A substantive body of literature describes judicial decision-making as a political behavior influenced by various political factors, including judges' policy preferences and party affiliation \cite{lax2011new, nagel1961political, schubert1958study, zorn2010ideological}. Our analysis focuses specifically on the political ideology of the US Supreme Court, given its role as the ultimate reviewer of all federal cases and state cases related to federal law. This central position of the US Supreme Court in the national legal hierarchy makes regular judges care about the preferences and inclinations of US Supreme Court justices \cite{baum2017judges}.

We propose that increasing ideological polarization among US Supreme Court justices, a phenomenon long documented in the literature \cite{devins2017split,hasen2019polarization}, has contributed to the decrease in case law disruption. As lower-court judges' decisions are ultimately subject to review by the Supreme Court, high polarization among the Supreme Court justices creates ambiguity in the institutional environment where their decisions are eventually evaluated. This ambiguity may lead lower-court judges to exhibit greater caution, thereby reducing their willingness to issue disruptive rulings. Therefore, we propose that the extent of ideological polarization among US Supreme Court justices in a given year is negatively related to the disruptiveness of judicial decision made in the same year. We measure political polarization as the standard deviation of justices' Martin-Quinn scores, a widely accepted metric for quantifying the ideological positions of US Supreme Court justices \cite{martin2002dynamic} (see \textbf{Methods} for details).

\begin{table*}[h!]
\centering
\begin{tabular}{ p{6.5cm} c c c c }
    \hline
    & \multicolumn{2}{c}{Model 1} & \multicolumn{2}{c}{Model 2}\\
    & \multicolumn{2}{c}{(1800-2000)} & \multicolumn{2}{c}{(1937-2000)}\\
    \hline
    Log number of cumulative cases in the jurisdiction by the year & -0.11** & (0.03) & -0.01 & (0.04) \\
    Court level (Supreme = 1) & 0.04** & (0.01) & 0.3 & (0.02) \\
    US Supreme Court political ideology (conservative) & & & -0.04* & (0.02) \\
    US Supreme Court political ideology dispersion & & & -0.10** & (0.03) \\
    \hline
    Year fixed effect & \multicolumn{2}{c}{Yes} & \multicolumn{2}{c}{Yes} \\
    Jurisdiction fixed effect & \multicolumn{2}{c}{Yes} & \multicolumn{2}{c}{Yes} \\
    N & \multicolumn{2}{c}{3,879,331} & \multicolumn{2}{c}{2,436,676} \\
    \hline
\end{tabular}
\caption{\textbf{The effect of volume and politics on case law disruption} The regression analysis shows the role of case law volume and political ideology in case law disruption. Standard errors (in parentheses) are clustered at the jurisdiction level. $* p < 0.05, ** p<0.01, *** p<0.001$ }
\label{table:regression}
\end{table*}

To test these two propositions about the role of case law volume and ideological polarization in the growing inertia, we conducted regression analyses. Table~\ref{table:regression} shows the results of two regression models: Model 1 (without ideology variables) and Model 2 (with ideology variables). Note that the two models have different time windows as the Martin-Quinn score is only available after 1937. Both models included fixed effects for years and jurisdictions. We encompassed 51 jurisdictions: the 50 US states and the federal jurisdiction.

The result shows that the cumulative number of cases in a jurisdiction by year has a negative effect on disruptiveness in Model 1, indicating that the mass of the system indeed increases its inertia. However, in Model 2, when we restrict the time window (1937-2000) and include ideology factors, the significance of the effect fades away, suggesting that the impact of case law volume may be context-dependent, or possibly diluted by political dynamics post-1937.

For the ideology variables, we included both the average and dispersion of the US Supreme Court justices' political ideology. First, the average political ideology of justices (conservatism) turned out to be negatively associated with case law disruption, indicating that the Supreme Court's conservatism leads to less disruptive rulings. Second, consistent with our proposition, greater ideological polarization among justices had a negative effect on disruptiveness. This means that when the Supreme Court is more ideologically divided, judges tend to issue less disruptive rulings.

These results demonstrate that both increasing legalization and growing ideological polarization within the highest court can suppress disruption in legal evolution. The rapid proliferation of precedents makes it increasingly difficult for judges to introduce disruptive opinions into an ever more complex legal landscape. At the same time, political divides within the highest court reinforce judicial adherence to the status quo, leading to gridlock in legal evolution.

\section*{Judicial hierarchy and issue areas}

\begin{figure*}[h!]
\centering

\includegraphics[width=.99\textwidth]{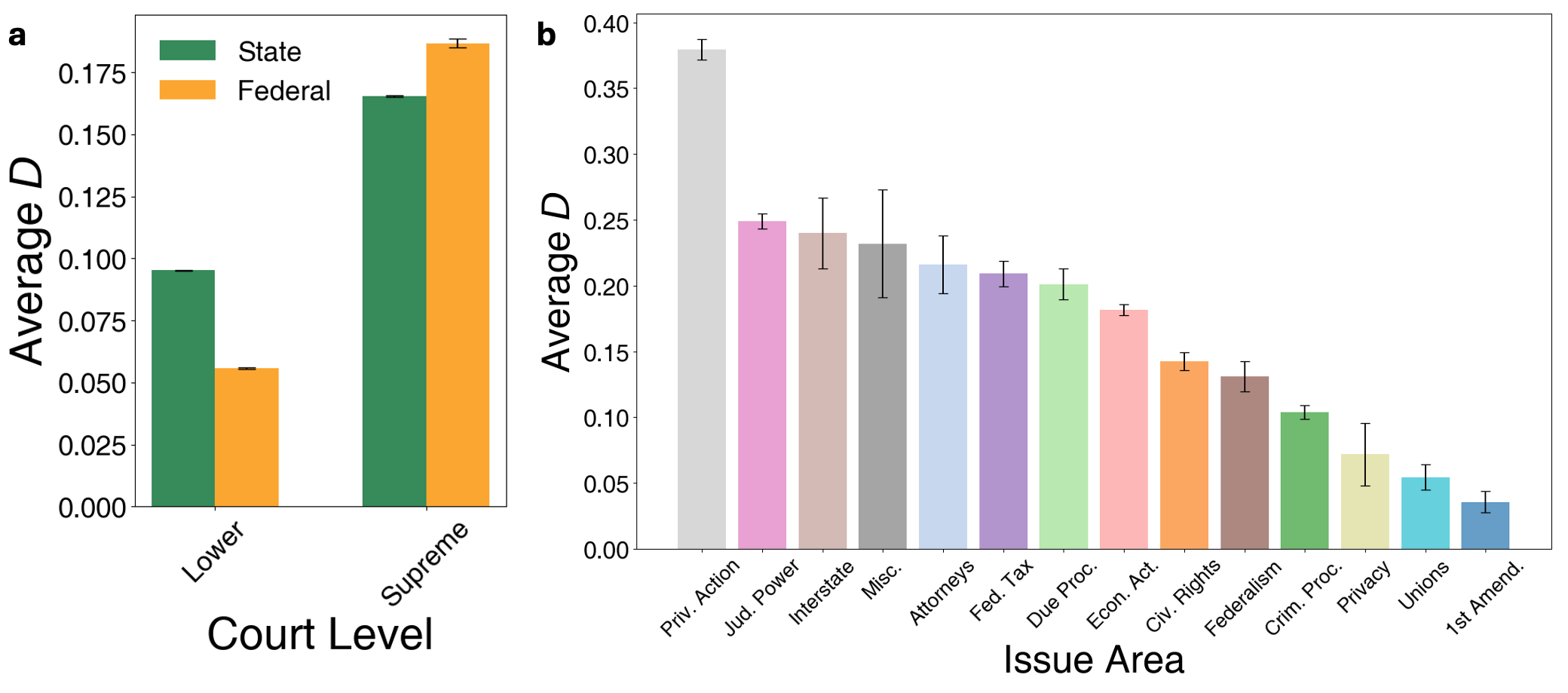}
\caption{\textbf{(a) Disruption index by court level: Lower and Supreme \& (b) Disruption index by issue area (US Supreme Court)} \textbf{a} Judicial decisions issued by higher-level courts are more disruptive in both federal and state jurisdictions. \textbf{b} The average $D$ of US Supreme Court cases varies significantly across issue areas. \newline Abbreviation: Priv. Action, Private Action; Jud. Power, Judicial Power; Interstate, Interstate Relations; Misc., Miscellaneous; Fed. Tax, Federal Taxation; Due Proc., Due Process; Econ. Act., Economic Activity; Civ. Rights, Civil Rights; Crim. Proc., Criminal Procedure; 1st Amend., 1st Amendment}
\label{fig:court_level_issue_area}
\end{figure*}

Beyond the two macro-level forces, i.e., legal proliferation and ideological polarization, which shape the institutional environment for all cases across the jurisdiction, case-specific factors, such as the authority of the ruling court and issue area, may also influence the likelihood of an individual decision disrupting the existing legal landscape. It is likely no coincidence that highly disruptive cases in our analysis, \textit{Brown} and \textit{Strauder}, were decided by the US Supreme Court, the pinnacle of the judicial hierarchy.

Hence, this section first examines how individual decisions' disruptiveness is shaped by the ruling court's authority. In the US, courts operate within a hierarchical structure, in which higher courts have the authority to review and affirm or overturn rulings made by lower courts. This hierarchy imposes structural constraints on lower courts, limiting the potential impact of their rulings in future cases even if those rulings are doctrinally disruptive. In contrast, supreme courts, which occupy the apex of their respective jurisdictions, are not subject to appellate review. Their institutional autonomy gives them greater latitude to issue transformative decisions that depart from precedent and reshape the legal landscape.

Figure~\ref{fig:court_level_issue_area}\textbf{a} shows that both federal and state supreme courts issue significantly more disruptive rulings than their lower-level counterparts. On average, decisions from the US Supreme Court are more than three times as disruptive as those from federal district or circuit courts. This pattern is echoed in the time trend shown in Figure~\ref{fig:d_by_year}, where the dotted lines (average disruptiveness of supreme court cases) deviate upward from the solid lines (average disruptiveness of all cases) in recent years. Notably, the US Supreme Court exhibited several surges in disruptiveness during the late 20th century. This suggests that especially during periods of overall decline in disruption, the nation's highest court may play a key role in breaking precedent and responding to evolving societal needs.

In addition to the variation introduced by court hierarchy, the issue domain of a case may also influence its potential for disruption. Some areas of law may have greater interpretive flexibility, leading to more precedent-breaking decisions, while others are more tightly constrained by constitutional mandates or deeply entrenched legal reasoning. To examine how issue areas shape disruptiveness, we match our disruption index to a dataset of US Supreme Court cases categorized by legal issue area~\cite{supremecourtdatabase}.

Figure~\ref{fig:court_level_issue_area}\textbf{b} shows substantial variation in the average disruptiveness of rulings across domains. For example, Private Action cases exhibit the highest average disruption ($D = 0.38$), followed by Judicial Power and Interstate Relations. In contrast, First Amendment cases show notably low levels of disruptiveness ($D = 0.04$), suggesting greater doctrinal stability in areas about foundational principles of the consitution.

If issue areas vary in their propensity to generate disruptive decisions, changes in the US Supreme Court’s issue composition over time may further explain the overall decline in disruptiveness. Indeed, our analysis reveals significant changes in the Court's attention over time (see Figure S3). In the 19th century, the docket was dominated by areas with a higher tendency to disrupt, such as Private Action. In contrast, issue areas with relatively lower disruptiveness, such as Unions, Civil Rights, and Criminal Procedure, have gained prominence in more recent decades. This pattern indicates that historical shifts in societal issues, each with different propensities for legal disruption, could be a factor that has contributed to the overall decline in disruption. While it is beyond the scope of this paper to conduct a deep qualitative examination of the historical context to further clarify the role of issue areas in the macroscopic trend, the analysis presented here offers a clue for better understanding the observed decline in disruptiveness over time.

\section*{Discussion}

We provide empirical evidence of a long-term decline in the legal system’s tendency to break precedent, based on nearly four million judicial opinions issued between 1800 and 2000. This finding illuminates the growing tension between historical continuity and adaptive responsiveness as legal systems age. In an era of rapid technological and societal transformation, a system structurally limited in its capacity to adapt may risk becoming misaligned with the world it is meant to govern. Understanding the sources of this legal inertia is therefore essential for sustaining institutional relevance over time.

Our analysis ultimately shows that law functions as a complex adaptive system, shaped by structural scale, institutional memory, and hierarchical structure. First, structural scale limits the system's adaptive capacity. As the body of case law expands on a superlinear scale relative to population, the cost of legal navigation grows disproportionately. Judges face increasing difficulty reconfiguring prior rulings, which results in greater reliance on established interpretive pathways. In this light, legal inertia is not just attributed to ideological conservatism but perhaps to an emergent property of scale.

Second, institutional memory, encoded in precedent, reinforces path dependence. It provides consistency across time and stabilizes legal doctrine, but it can also leave outdated norms. That is, precedent serves not only as a repository of legal reasoning but also as a structural constraint on doctrinal innovation.

Third, hierarchical structure further shapes the system’s capacity for disruption. Lower courts, bound by institutional deference, largely reproduce doctrinal continuity. In contrast, high-authority courts, particularly the US Supreme Court, retain the discretion to redirect legal trajectories. Landmark disruptive rulings from the US Supreme Court, such as \textit{Strauder v. West Virginia} (1879), \textit{Brown v. Board of Education} (1954), and \textit{Dobbs v. Jackson Women’s Health Organization} (2022), illustrate how top-down interventions can reorient the legal landscape. While this functional devision of labor between high- and low-authority courts allows the overall system to maintain its continuity while incorporating innovations, it still raises critical questions about system design: Who should have the authority to determine which innovations are made? And what risks arise when responsiveness is reserved for the few?

Our study has several limitations. First, the disruption metric does not directly speak to how the substantive content of legal doctrine has changed. This structural property has strength in large-scale comparative analysis, but inherently entails a lack of specific context. While we illustrate its implications through example cases like \textit{Plessy} and \textit{Brown}, future work could trace the evolution of legal content more explicitly. Moreover, valid critiques concerning statistical artifacts associated with the disruption index warrant consideration. Although we address some of these through robustness checks, a comprehensive empirical understanding of legal disruption remains an open challenge. All of this points to the need for developing diverse measures for legal disruption. 

Second, our analysis identifies correlations rather than causal effects. Although we control for jurisdiction and year fixed effects in the regression models, the results remain observational. Unmeasured confounders may influence both legal outcomes and institutional structure. Future work using quasi-experimental designs or natural experiments could help isolate the drivers of declining disruptiveness.

Despite these limitations, our study contributes by situating the study of law within a broader framework of institutional evolution and innovation. This analytical lens posits legal reasoning not merely as an individual act of interpretation or deliberation, but as an activity deeply embedded within a recursive structure shaped by historical accumulation and power dynamics. Exogenous shocks, such as the Civil War followed by the Reconstruction Era, continue to generate moments of doctrinal change. But in the absence of such shocks, the system exhibits a strong tendency toward continuity. This perspective helps explain both the decline in legal disruption and the conditions under which structural transformation nonetheless becomes possible.

\section*{Materials and Methods}
\subsection*{Data}
This study uses a public dataset on US judicial opinions from the Caselaw Access Project (retrieved on November 9, 2023)~\cite{caselawproject}. The dataset provides a comprehensive digitized collection of over six million state and federal cases published in the US over more than 360 years, from 1658 to 2020. It includes metadata on cases, such as the ruling court, year of issuance, and citation relationships, which we use to construct the citation network and calculate the disruption index.

For this paper, we focus on judicial opinions published between 1800 and 2000. We begin our analysis in 1800, as the US federal government was established under the Constitution by the late 18th century. We set the upper limit at 2000 because calculating a meaningful disruption index requires enough time for future cases to cite a given case. Since cases are less likely to be cited after about 20 years \cite{black2013citation, whalen2017common}, we use a 20-year window between the end of our analysis period and the most recent cases in the dataset. This brings the total number of cases used in our analysis to 3,976,542.

\subsection*{Disruption index in case law}
We use the disruptive index, $D$, to capture the disruptiveness of each case, ranging from -1 (consolidating) to 1 (disruptive)~\cite{funk2017dynamic}. For each case, cases that cite either the focal one or its predecessors are grouped into three categories: (1) cases that cite only the focal case but none of its predecessors (type $i$), (2) cases that cite both the focal case and its predecessors (type $j$), and (3) cases that cite the predecessors but not the focal case (type $k$). The disruption index, $D$, is calculated as follows:

\[
D  = \frac{n_i-n_j}{n_i+n_j+n_k}
\]

Figure~\ref{fig:schematic} illustrates how this relative precedential power is captured in Eq. \ref{eq:d_index}, where $n_i$, $n_j$, and $n_k$ represent three different groups of cases related to the focal case: $n_i$ (green) counts those that cite only the focal case, $n_j$ (blue) counts those that cite both the focal case and its predecessors, and $n_k$ (grey) counts those referencing only the original cases that the focal case references but not the focal case itself. A higher proportion of cases that exclusively cite the focal case ($n_i$) indicates a more disruptive impact introduced by the focal case law. As a result, $D_{Brown} = 0.33  \gg D_{Plessy} = 0.03$ by nearly tenfold more disruptive effect on the legal landscape.

To ensure the robustness of our analysis, we tested several variations of the disruption measure. Although, in principle, a court decision's precedential power persists indefinitely unless explicitly overturned, empirical research shows that most decisions experience a decline in citation frequency after roughly 20 years \cite{black2013citation, whalen2017common}. To account for this, we computed an alternative measure, $D_{20}$, which includes only forward citations made within 20 years of the decision. The declining trend in disruptiveness persisted under this specification (see Figure S1).

We also tested another alternative, $D_{nok}$ \cite{bornmann2020disruption, wu2019solo}, which has been used in the literature to address concerns that citation inflation might artificially drive the observed decrease in the disruption index \cite{park2023papers}. $D_{nok}$ mitigates this issue by excluding $n_k$ from the denominator, which is deemed sensitive to the number of references cited by a focal work. We found that the decreasing pattern remained consistent (see Figure S2). Indeed, the decline appears to be even steeper when the alternative measure is applied, which suggests that citation inflation is not a significant issue in case law. These two additional analyses indicate that the decrease in case law disruptiveness is not likely a technical artifact, but rather a real phenomenon.

\subsection*{Scaling analysis of case law}
We conduct scaling analysis~\cite{bettencourt2007growth, youn2016scaling, west2018scale, west1997general} to see how the number of cases ($Y$) changes as the population size ($N$) increases. This analysis allows us to understand the underlying patterns and trends in legal development in relation to demographic changes over an extended historical period. The relationship can be expressed using the following equation:

\[
Y = Y_0 N^\beta
\]

where $Y$ represents the number of cases, $N$ denotes the population size, $Y_0$ is a normalization constant, and $\beta$ is the scaling exponent that characterizes the relationship between population size and the number of cases. Values of $\beta$ exceeding 1 indicate a faster-than-linear increase in the number of cases with population size, while values of $\beta$ below 1 indicate a slower-than-linear increase.

\subsection*{Political ideology index}
We use the Martin-Quinn Scores~\cite{martin2002dynamic} to assess the political ideology of individual US Justices. These scores quantify justices' positions on a liberal-conservative spectrum based on their voting behavior in legal cases. Negative scores indicate a liberal leaning, while positive scores indicate a conservative leaning. A score of zero suggests a moderate or centrist position. These scores are dynamic and can change over time, reflecting shifts in a justice's voting behavior. We calculate the average of individual US Justices' Martin-Quinn scores in a given year to obtain the overall political leaning of the US Supreme Court, and use the standard deviation of these scores to measure the level of polarization within the US Supreme Court.

Given that the political ideology data span from 1937, our regression analysis with variables about ideology is restricted to 1937-2000. The models are estimated using ordinary least squares (OLS) regressions.

\subsection*{Disruption index by issue area}
We use the Supreme Court Database~\cite{supremecourtdatabase}, which links US Supreme Court decisions to various case-level characteristics, to identify the issue area of each case. We matched the issue area classifications from this database to the corresponding US Supreme Court cases in the Caselaw Access Project dataset, which we used to compute the disruption index, by aligning their citation formats. This allows us to calculate the average disruption index for each issue area.

\section*{Acknowledgements}
The authors would like to acknowledge the support of the National Science Foundation Grant Award Number 2133863. H.Y. and J.Y. acknowledge Global Humanities and Social Sciences Convergence Research Program through the National Research Foundation of Korea (NRF), funded by the Ministry of Education (2024S1A5C3A02042671). H.Y. acknowledges the support from the Institute of Management Research at Seoul National University and the Emergent Political Economies grant from the Omidyar Network through Santa Fe Institute. S.L. thanks Klaus Weber and Brian Uzzi for reading the first draft of this study and providing insightful comments and references. J.Y. thanks B.K. Kim for insightful discussion and comments.

\section*{Author Contributions}

All authors designed the study, performed data analysis, interpreted the results, wrote the manuscript, and approved the final version of the manuscript. 

\section*{Additional Information}
Supporting Information is available for this paper. Correspondence and requests for materials should be addressed to Dr. Youn.

\section*{Data availability}
Case law data are available at Caselaw Access Project, \url{https://case.law/}.
 
 \section*{Code Availability}
The code for this analysis is available upon request.

\bibliographystyle{naturemag}
\bibliography{main}

\end{document}